\documentclass[aps,prl,reprint,superscriptaddress]{revtex4-1}

\usepackage{hyperref}
\usepackage{graphicx}
\usepackage{enumerate}
\hypersetup{
    pdfnewwindow=true,      
    colorlinks=true,       
    linkcolor=blue,          
    citecolor=blue,        
    filecolor=blue,      
    urlcolor=blue           
}

\usepackage{subfigure}
\usepackage{amsmath}
\usepackage{amssymb}
\usepackage{amsfonts}

\usepackage{color}

\providecommand{\apj}[0]{ApJ}

\providecommand{\nat}[0]{Nature}

\providecommand{\prd}{PRD}

\begin{document}

\title{Beyond the Horizon Distance: LIGO-Virgo can Boost Gravitational Wave \\ Detection Rates by Exploiting the Mass Distribution of Neutron Stars}

\author{I. Bartos}
\email{ibartos@phys.columbia.edu}
\affiliation{Department of Physics, Columbia University, New York, NY 10027, USA}
\affiliation{Columbia Astrophysics Laboratory, Columbia University, New York, NY 10027, USA}
\author{S. M\'arka}
\affiliation{Department of Physics, Columbia University, New York, NY 10027, USA}
\affiliation{Columbia Astrophysics Laboratory, Columbia University, New York, NY 10027, USA}

\begin{abstract}
The masses of neutron stars in neutron star binaries are observed to fall in a narrow mass range around $\sim 1.33$\,M$_{\odot}$.
We explore the advantage of focusing on this region of the parameter space in gravitational wave searches. We find that an all-sky
(externally triggered) search with optimally reduced template bank is expected to detect $14\%$ ($61\%$) more binary mergers than
without the reduction. A reduced template bank can also represent significant improvement in technical cost. We also develop a
more detailed search method using binary mass distribution, and find similar sensitivity increase to that due to the reduced template bank.
\end{abstract}


\maketitle

Binary neutron star (NS) mergers represent one of the most promising source type for gravitational wave
(GW) detection \cite{2013arXiv1304.0670L,LIGOwhitepaper}. With the recent onset of observations with the
advanced LIGO detectors \cite{2015CQGra..32g4001T} and with advanced Virgo and KAGRA
in the near future \cite{2015CQGra..32b4001A,PhysRevD.88.043007}, the first detections are expected
within the next few years \cite{2013arXiv1304.0670L}.

Current searches for compact binary mergers aim to cover virtually the full plausible binary parameter space \cite{LIGOwhitepaper}.
The strategy is also motivated by the fact that search sensitivity is considered to be only weakly dependent on the extent of the
covered parameter space \cite{1998PhRvD..57.4535F}.

The primary search method for GWs from compact binary mergers is the use of matched filters \cite{1998PhRvD..57.4535F,LIGOwhitepaper}.
This method correlates the known signal waveform, called \emph{template}, with the data to identify a GW signal.
For given binary parameters, the signal waveform can be calculated to high precision.
The binary parameter space is then covered by using a large number of templates, called the \emph{template bank},
such that the search sensitivity is sufficiently close to optimal within the considered parameter space (e.g., \cite{2012PhRvD..86h4017B}).

The properties of NS binaries are increasingly constrained due to the growing number of observed binaries.
These observations suggest that the mass of NS within NS binaries is within a surprisingly small
range around $\sim 1.33$\,M$_{\odot}$, much smaller than the allowed NS mass range from $\sim 1$\,M$_\odot$
to $\lesssim 3$\,M$_\odot$, or even the mass range of NS in NS-white dwarf binaries \cite{2013ApJ...778...66K}.

In this paper we investigate the effect of the physically motivated reduction of the template bank on search sensitivity.
We consider (i) blind, so-called \emph{all-sky} searches (e.g., \cite{2012PhRvD..85h2002A}) in which only GW data is utilized,
as well as (ii) so-called \emph{externally triggered} searches (e.g., \cite{2008ApJ...681.1419A}), in which the electromagnetic
or other detection of the binary merger aids the GW search. After discussing the dependence of the search sensitivity on the size
of the template bank, we optimize sensitivity as a function of the confidence region of the NS masses in the binary for
the two search strategies. Finally, we develop a more detailed search that incorporates a ranking statistic for the templates based
on the expected mass distribution of NS binaries. We calculate the advantage of such search over a baseline search that
uniformly weights templates.

\begin{figure}
\begin{center}
\resizebox{0.48\textwidth}{!}{\includegraphics{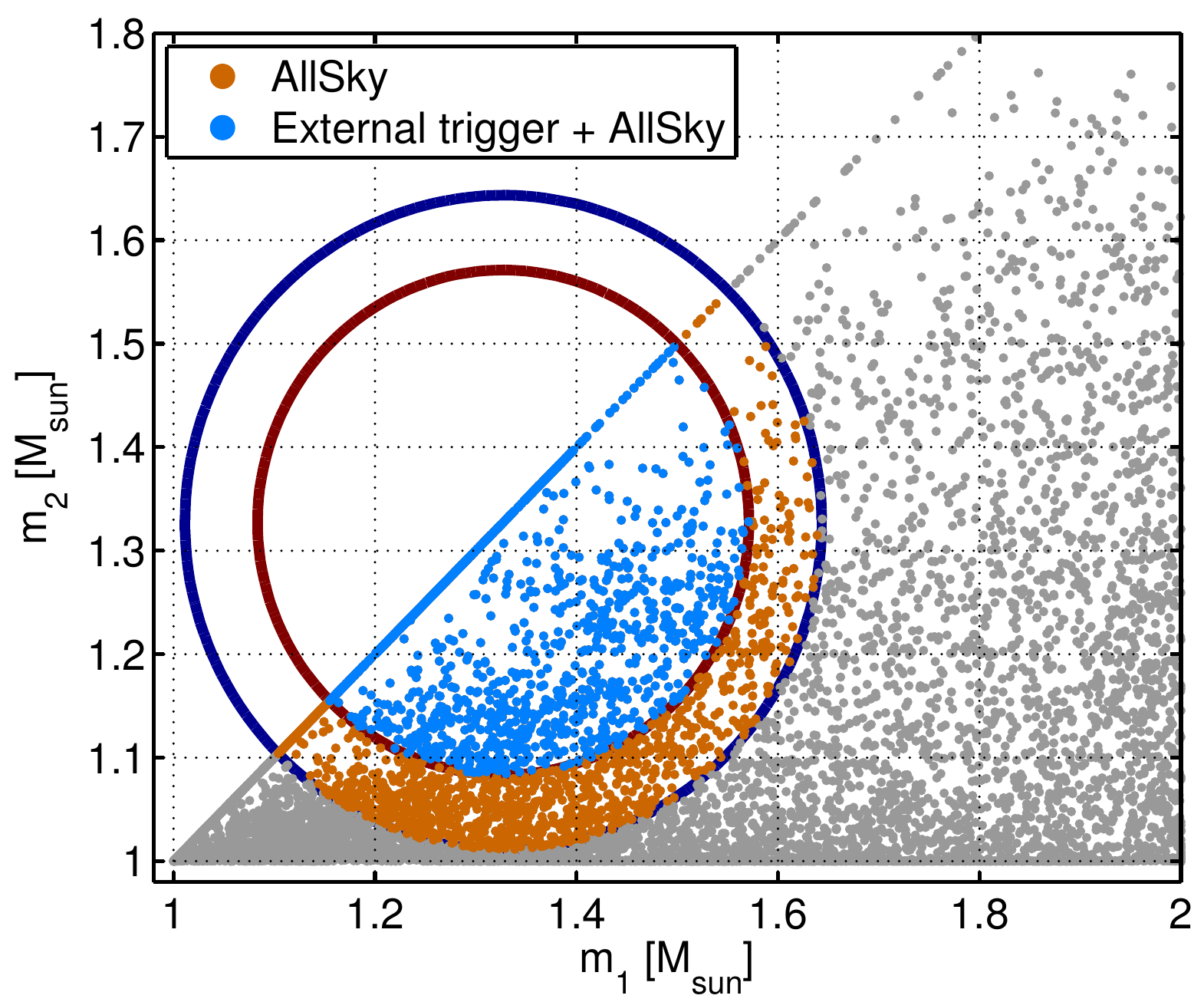}}
\end{center}
\caption{Distribution of GW binary templates in the parameter space of the masses of the two NSs in the binary. 
Circles: $92\%$ (smaller) and $99\%$ (bigger) mass confidence regions based on the empirical binary NS mass distribution
of \cite{2013ApJ...778...66K}. Templates within the bigger circle (orange+blue) correspond to the optimally reduced all-sky search. 
Templates within the smaller circle (blue) correspond to the optimally reduced externally triggered search.
}
\label{figure:TemplateCut}
\end{figure}

\vspace{0.3cm}
\paragraph{Sensitivity dependence on template bank. ---}
Let $\widetilde{h}(f)$ be the known gravitational waveform and $S_{\rm f}(f)$ be the power spectral density of strain noise in the detector.
The signal-to-noise ratio (SNR) of a matched-filter-based search is \cite{1998PhRvD..57.4535F}
\begin{equation}
\rho^2 = \left(\frac{S}{N}\right)_{\rm matched}^2 = 4 \int_{0}^{\infty}\frac{|\widetilde{h}(f)|^{2}}{S_{\rm f}(f)} df.
\end{equation}
Assuming the simple case of a single GW detector with stationary Gaussian noise, a detection can be claimed
when the matched filter SNR exceeds a threshold $\rho_{\rm th}$.
This threshold depends on, among others,
the trial factor $\mathcal{N}_{\rm trial}$ associated with the search. For Gaussian noise, the
threshold satisfies \cite{1998PhRvD..57.4535F}
\begin{equation}
{\rm erfc}(\rho_{\rm th} / \sqrt{2}) \approx \frac{\mbox{FAP}}{\mathcal{N}_{\rm trial}},
\label{eq:threshold}
\end{equation}
where FAP is the false alarm probability of the search. LIGO typically considers $\rho_{\rm th} = 8$ in a single 
detector as a detection threshold \cite{2013arXiv1304.0670L}. We adopt this value in the following.

The trial factor $\mathcal{N}_{\rm trial}$ depends on the number $\mathcal{N}_{\rm templates}$ of templates used,
along with the number $\mathcal{N}_{\rm t}$ of independent starting times that are considered \cite{1998PhRvD..57.4535F},
and for analyses with multiple detectors, the number $\mathcal{N}_{\rm \Omega}$ of independent source directions the search
considers. Using the fact that $\mathcal{N}_{\rm t} \propto t_{\rm obs}$, where $t_{\rm obs}$ is the observation duration,
and $\mathcal{N}_{\rm \Omega} \propto \Omega$, where $\Omega$ is the allowed sky region, the trial factor satisfies
\begin{equation}
\mathcal{N}_{\rm trial} \propto \mathcal{N}_{\rm templates} \, t_{\rm obs} \, \Omega.
\label{eq:trial}
\end{equation}
We see that changing the size of the template bank changes $\mathcal{N}_{\rm trial}$, which in turn will
change the search sensitivity through $\rho_{\rm th}$.

Besides the effect of the template bank, we also see that $t_{\rm obs}$ and $\Omega$ similarly affect search sensitivity,
which is relevant for externally triggered searches.

We now derive a formula to quantify the change in sensitivity due to varying the search parameters. Let our baseline
search have $\mathcal{N}_{\rm templates}^{(0)}$ templates, $t_{\rm obs}^{(0)}$ observation time, include $\Omega^{(0)}$
sky area, and have $\rho_0$ detection threshold. We want to determine the new threshold $\rho_1$ if we change the
parameters to $\mathcal{N}_{\rm templates}^{(1)}$, $t_{\rm obs}^{(1)}$ and $\Omega^{(1)}$, respectively. We can use
Eq. \ref{eq:threshold}, which yields, to a good approximation,
$\rho_{\rm th} \approx \sqrt{2\ln(\mathcal{N}_{\rm trial} / \mbox{FAP})}$ \cite{1998PhRvD..57.4535F}.
With this, we obtain
\begin{equation}
\rho_1 \approx \sqrt{2 \ln \left[\exp\left({\frac{\rho_{0}^{2}}{2}}\right)\frac{\mathcal{N}_{\rm templates}^{(1)}t_{\rm obs}^{(1)}\Omega^{(1)}}{\mathcal{N}_{\rm templates}^{(0)}t_{\rm obs}^{(0)}\Omega^{(0)}}\right]}
\label{eq:rho}
\end{equation}

\vspace{0.3cm}
\paragraph{Utilizing neutron star mass distribution. ---}
The masses of NSs within observed NS binaries fall in a surprisingly small range.
While the maximum allowed NS mass is above 2\,M$_\odot$ \cite{2010Natur.467.1081D}, NS masses
in NS binaries are closely clustered around 1.33\,M$_\odot$. Kiziltan \emph{et al.} \cite{2013ApJ...778...66K}
use the observed masses and their uncertainties in a statistical model to find an empirical NS mass ($m$) distribution
\begin{equation}
P_{\rm ns}(m) = 2 \phi\left(\frac{m-\mu}{\sigma}\right)\Phi\left(\frac{(m-\mu)\alpha}{\sigma}\right),
\label{eq:pm}
\end{equation}
where $\phi(x)$ and $\Phi(x)$ are the standard normal density and cumulative density functions, respectively.
For NS binaries, Kiziltan \emph{et al.} find $\mu=1.33$, $\sigma = 0.11$ and $\alpha = -0.03$. We adopt this empirical
NS mass distribution in the following. We conservatively assume that the two NS masses in the binary are independent.

We utilize the NS mass probability distribution by using only those templates in a matched-filter-based
search that are more likely to be observed from astrophysical sources. This way, we can significantly reduce the size of the template bank
with modest reduction of the fraction $f_{\rm ns}$ of NS binaries whose masses are covered by the template bank.
Let $Y_{\rm ns}$ be the region in the 2D mass parameter space that is included in the analysis. 
For any $f_{\rm ns}$, we can define $Y_{\rm ns}$ such that (i) a template that is $\in Y_{\rm ns}$ is
at least as likely to correspond to a detected NS binary as any template that is $\notin Y_{\rm ns}$, and (ii) the fraction of NS binaries that fall
within $Y_{\rm ns}$ is $f_{\rm ns}$.


\vspace{0.3cm}
\paragraph{Sensitivity improvement. ---}
The most meaningful quantity to compare searches with is the detection rate $\mathcal{R}$, which can be written as
\begin{equation}
\mathcal{R} = \frac{4}{3}\pi \rho_{\rm th}^{-3} \, f_{\rm ns} \, f_{b}^{-1} \, f_{\rm gw}.
\label{eq:rate}
\end{equation}
Here, $f_b$ is the beaming factor of the emission corresponding to the external trigger, which reduces the number of observed sources \cite{2013CQGra..30l3001B};
for all-sky searches, we can take $f_b = 1$. For externally triggered searches, the direction of the external trigger
can be correlated with the weakly direction-dependent GW emission. For triggers such as GRBs, which are aligned with
the orbital axis of the binary merger, for $f_b \gg 1$ we can approximate $f_{\rm gw} \approx 1.5$ \cite{2013CQGra..30l3001B},
which is the ratio of the GW strain amplitude in the direction of the orbital axis compared to the directionally averaged strain amplitude.

To estimate the sensitivity improvement in different search scenarios, we adopt the template bank used by initial LIGO-Virgo in 
\cite{2012PhRvD..85h2002A}, which is representative of what will be used for advanced LIGO-Virgo searches \cite{LIGOwhitepaper} 
(see Fig. \ref{figure:TemplateCut} for the distribution of templates for the mass regime $\lesssim 2$\,M$_{\odot}$). The templates are
distributed such that the loss in SNR due to the discreteness of the template bank is less than $3\%$,
but otherwise cover the parameter space for masses $\geq1$\,M$_\odot$. To demonstrate the role of the size of the template
bank for binary NS searches, we focus on the part of the bank for which both masses in the binary are $<3$\,M$_\odot$.
This is the parameter range planned to be used for binary NS searches with advanced LIGO \cite{LIGOwhitepaper}. 
The part of the template bank below $3$\,M$_\odot$ includes $\sim 45,000$ templates. We denote
the detection rate of this baseline search with $\mathcal{R}_{\rm{allsky}}$.

\begin{figure}
\begin{center}
\resizebox{0.48\textwidth}{!}{\includegraphics{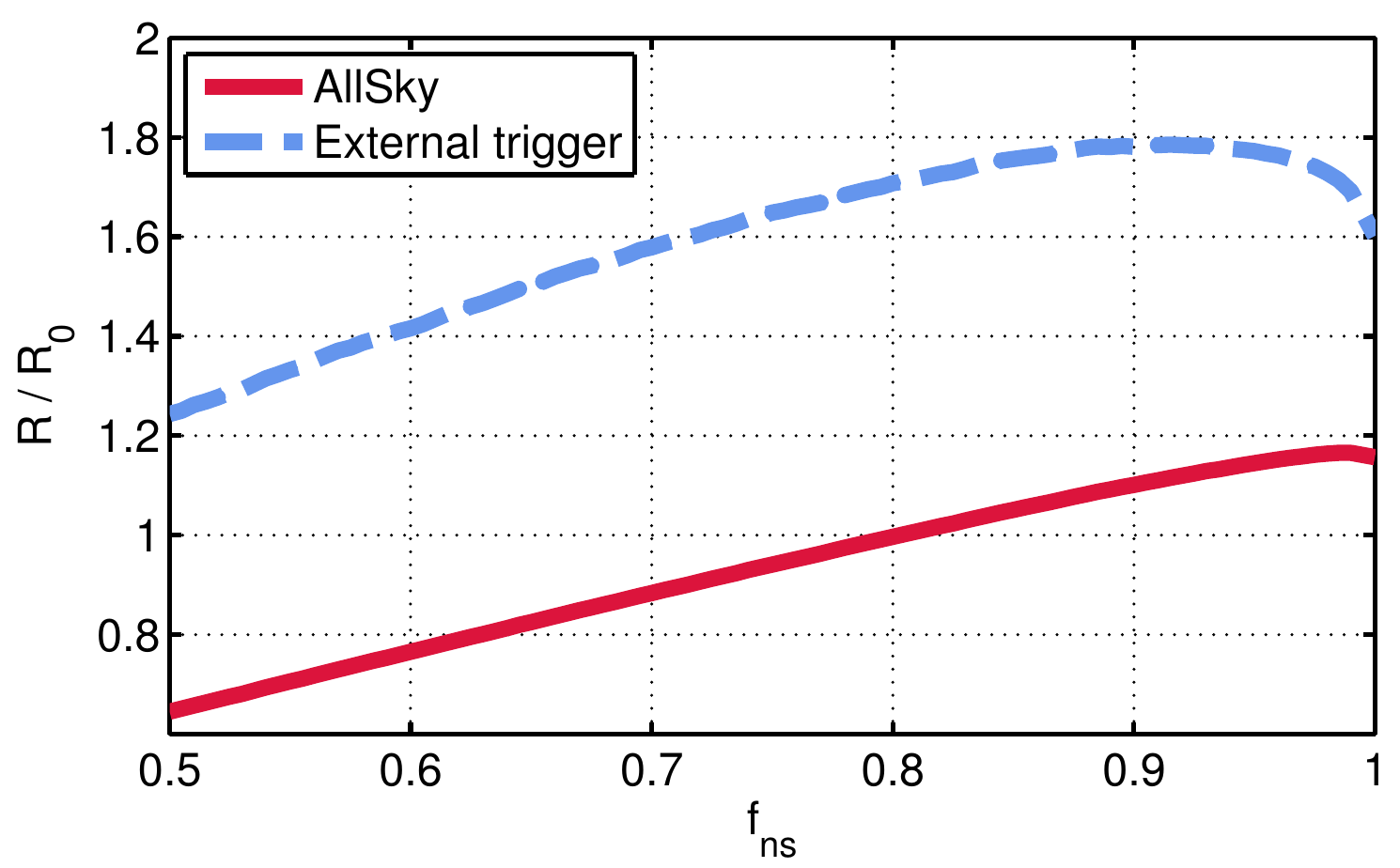}}
\end{center}
\caption{Improvement of detection rate using the constrained NS mass parameter space
over the baseline search with no constraints.
Results are shown for all-sky (solid) and externally triggered
(dashed) searches, as functions of the fraction $f_{\rm ns}$ of the NS mass parameter space included in the template bank.}
\label{figure:detectionrate}
\end{figure}

We first consider all-sky observations. We assume $t_{\rm obs} = 1$\,yr observation time, and $\Omega = 4\pi$. 
The baseline search, i.e. the one without taking advantage of the NS mass distribution, will use the
full template bank below $3$\,M$_\odot$. We take a detection threshold of $\rho_{\rm{allsky}} = 8$
\cite{2013arXiv1304.0670L}. Compared to this baseline, an all-sky search using a constrained binary NS parameter space
will yield detection threshold $\rho_{\rm{allsky,NS}} < \rho_{\rm{allsky}}$ due to the reduced template bank. The detection rate
for the constrained parameter space, normalized by the detection rate for the baseline all-sky search, is shown in Fig. \ref{figure:detectionrate}
as a function of the NS confidence level $f_{\rm ns}$. We select the $f_{\rm ns}$ value that
maximizes the detection rate $\mathcal{R}_{\rm{allsky,NS}}$. The corresponding value is $f_{\rm ns,allsky} = 0.99$,
with $\rho_{\rm{allsky,NS}} = 7.6$ and rate $\mathcal{R}_{\rm{allsky,NS}}/\mathcal{R}_{\rm{allsky}} = 1.14$.
The constrained parameter space and the corresponding template distribution are shown in Fig. \ref{figure:TemplateCut}.

We next look at externally triggered searches.
External triggers can significantly boost search sensitivity by decreasing the time window and sky area
in which GWs need to be searched for. Similarly to Chen and Holz \cite{2013PhRvL.111r1101C}, we consider
$t_{\rm obs} = 10$\,s, corresponding to the approximate GW search time window for one sufficiently nearby
GRB detected, and $\Omega = 100$\,deg$^{2}$, an approximate sky area corresponding to a known
signal direction, taking into account the directional precision of GW searches. For the baseline externally
triggered search, we do not take advantage of the NS mass distribution, and use the full
template bank below $3$\,M$_\odot$. Using Eq. \ref{eq:rho}, we find that the detection threshold of this
baseline search is $\rho_{\rm{extrig}} = 4.7$. We now obtain the detection rate improvement for the case
with constrained NS parameter space. Similarly to the all-sky case, we select the $f_{\rm ns}$
value that maximizes the detection rate $\mathcal{R}_{\rm{extrig,NS}}$. The corresponding value is $f_{\rm ns,extrig} = 0.92$,
with $\rho_{\rm{extrig,NS}} = 3.9$ and rate $\mathcal{R}_{\rm{extrig,NS}}/\mathcal{R}_{\rm{extrig}} = 1.61$,
a significant improvement over the baseline externally triggered case. 
The constrained parameter space and the corresponding template distribution are shown in Fig. \ref{figure:TemplateCut}.

We now explore the difference between detection rates for the all-sky and externally triggered cases. 
This difference depends on the beaming factor $f_b = [1-\cos(\theta)]^{-1}$, where $\theta$ is the opening angle. In general, 
externally triggered searches benefit from increased sensitivity, while all-sky searches are advantageous as only a fraction of
events will have external triggers due to beaming.
Using Eq. \ref{eq:rate}, we find that, for the baseline case,
$\mathcal{R}_{\rm{extrig}}/\mathcal{R}_{\rm{allsky}} = 0.25$ and 2.24 for $\theta=10^\circ$ and 30$^\circ$,
respectively. The two baseline searches are expected to detect the same number of binaries for $\theta = 20^\circ$.
For the constrained searches, $\mathcal{R}_{\rm{extrig,NS}}/\mathcal{R}_{\rm{allsky,NS}} = 0.36$
and 3.18 for $\theta=10^\circ$ and 30$^\circ$, respectively, significantly higher than for the baseline case.
The constrained externally triggered search is expected to do as well as the constrained all-sky search for
$\theta=17^\circ$. This means that for the typically expected short-GRB opening angles of $<10^\circ$
\cite{2012ApJ...756..189F}, the majority of the detected events will come from all-sky searches.

\vspace{0.3cm}
\paragraph{Search method incorporating NS mass distribution. ---}
To gain a more detailed picture of the advantages of utilizing the binary NS mass distribution in GW searches,
we calculate the increased sensitivity of a likelihood-ratio test that incorporates mass information
(for likelihood-based methods and their benefits, see \cite{2010PhRvD..82j2001A,2012PhRvD..85h9903A,2012PhRvD..85l2008B,2012PhRvD..85l2009B,2014PhRvD..89f2002D,2015arXiv150404632C}). We define our test statistic as
\begin{equation}
\mathcal{L} = \frac{P(m_1,m_2|\mbox{signal})P(\rho | m_1,m_2,\mbox{signal})}{P(m_1,m_2|\mbox{noise})P(\rho | m_1,m_2,\mbox{noise})}
\end{equation}
where $m_1$ and $m_2$ are the masses of the two NS in the binary, and $P(\cdot|\mbox{signal}/\mbox{noise})$
is the conditional probability for the signal/noise hypothesis. Here, $P(m_1,m_2|\cdot)$ is the probability
corresponding to the template with masses $m_1$ and $m_2$, i.e. it is \emph{not} a probability density.


For simplicity, we assume $P(m_1,m_2|\mbox{noise}) = \mbox{const}$, and that the probability distribution of $\rho$ is mass independent:
$P(\rho | m_1,m_2,\mbox{noise}) = P(\rho | \mbox{noise})$ and $P(\rho | m_1,m_2,\mbox{signal}) = P(\rho | \mbox{signal})$. Note that
$P(m_1,m_2|\cdot)$ considered here is the probability of a specific template defined by $m_1$ and $m_2$, i.e. not probability density.

Adopting a uniform binary NS distribution in the local universe yields
$P(\rho | \mbox{signal}) \propto \rho^{-4}$. Considering a Gaussian noise model, we get
$P(\rho | \mbox{noise}) \propto \exp(-\rho^2 / 2)$. With these assumptions, we arrive at
\begin{equation}
\mathcal{L} = P(m_1,m_2|\mbox{signal}) \, \rho^{-4} \, e^{\rho^2/2}.
\label{eq:TS}
\end{equation}
where we omitted a constant factor that is irrelevant for the statistic.

We consider three different models for $P(m_1,m_2|\mbox{signal})$. The baseline model that assigns no weight to the
specific $\{m_1,m_2\}$ values: $P(m_1,m_2|\mbox{signal})|_{\rm baseline} = const.$ A model with a cutoff in the allowed
mass parameter space, in line with our discussion above; for this model, $P(m_1,m_2|\mbox{signal})|_{\rm cutoff} = \mbox{const.}$
for $[(m_{1}-\mu)^{2}+(m_{2}-\mu)^{2}]^{1/2} < 3\sigma$ and $=0$ otherwise. In the third, weighted model, we fully
make use of the expected mass distribution by assigning the appropriate weight from Eq. \ref{eq:pm}:
$P(m_1,m_2|\mbox{signal})|_{\rm weighted} = P_{\rm ns}(m_1)P_{\rm ns}(m_2)$ (see e.g., \cite{2015arXiv150404632C} for context and motivation of weighted signal priors).

We are interested in calculating the sensitivity of the different models as functions of FAP. 
For this, we define FAP as the probability of obtaining $\mathcal{L}$
greater than a threshold value $\mathcal{L}_{\rm th}$ during the observation period for
any of the templates: $\mbox{FAP} \equiv P_{\rm obs}(\mathcal{L} \geq \mathcal{L}_{\rm th} | \mbox{noise})$.
Instead of $\mathcal{L}_{\rm th}$, we can also equivalently specify the threshold as a SNR 
threshold $\rho_{\rm th}$ using Eq. \ref{eq:TS}.
\begin{equation}
\mbox{FAP} \propto \sum_{i} \int_{\rho_{{\rm th},i}}^{\infty}P(\rho | \mbox{noise}) d\rho = \sum_{i}\left[1 - \Phi(\rho_{{\rm th},i})\right]
\label{eq:FAP}
\end{equation}
where $\rho_{{\rm th},i}$ is the threshold value of the $i^{\rm th}$ template such that
$\mathcal{L}(m_{1,i}, m_{2,i}, \rho_{{\rm th},i}) = \mathcal{L}_{\rm th}$, the sum is over all
templates in the bank, and $\Phi$ is the cumulative distribution function of the standard normal distribution.

\begin{figure}
\begin{center}
\resizebox{0.48\textwidth}{!}{\includegraphics{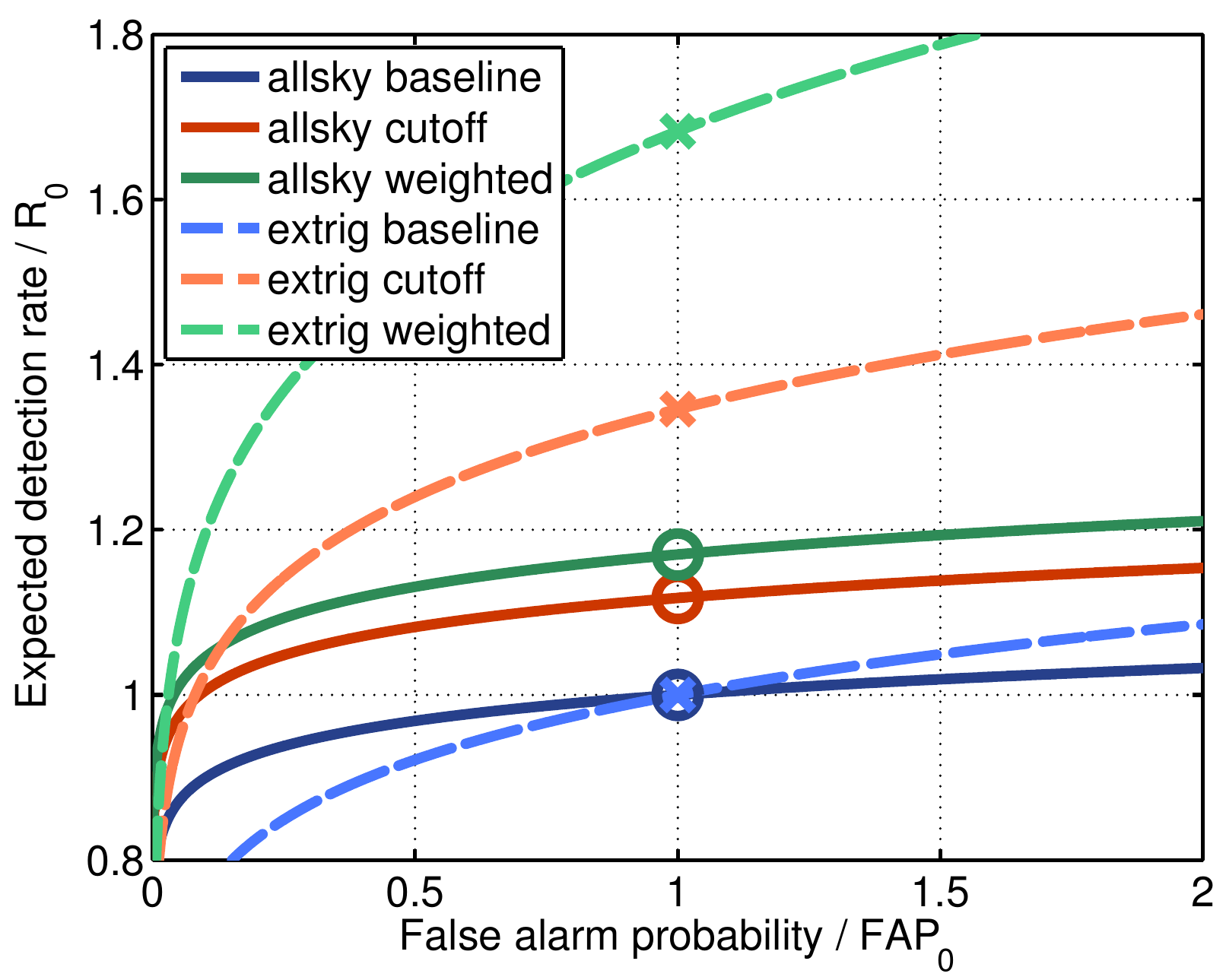}}
\end{center}
\caption{Expected detection rate as a function of the false alarm rate for different all-sky
and externally triggered search models using the likelihood-ratio test statistic, normalized to the baseline method. 
}
\label{figure:ROC}
\end{figure}

To calculate the search sensitivity as a function of FAP, we calculate the expected rate of
detected NS binaries. We consider a binary detected if its SNR is $>\rho_{\rm th}(m_1,m_2)$, where we note that the threshold depends on the binary
masses. For uniform source distribution, the expected rate of detected binaries is
$\propto \rho_{\rm th}^{-3}$. For simplicity, we neglect the effect of decreased
sensitivity due to binary masses being off of templates, and consider that the template with
the closest masses to those of the binary will be used. We define distance with respect to
mass with $(\Delta m_{1}^2+\Delta m_{2}^2)^{1/2}$. Using this definition, for template $i$
we determine the area $A_{i}$ within the mass parameter space that is closest to template $i$.
Assuming that, for a given binary merger, its SNR and the closest template
can be accurate recovered, the total detected rate can be expressed as
\begin{equation}
\mathcal{R} \propto \sum_{i} \rho_{{\rm th},i}^{-3} \, P(m_{1,i},m_{2,i}|\mbox{signal})\,A_{i}
\label{eq:R}
\end{equation}
Using Eqs. \ref{eq:FAP} and \ref{eq:R}, we can calculate the expected detection rate as a
function of FAP for each of the considered search models. The results are shown in Fig. \ref{figure:ROC}.

To compare the sensitivity of different search models, we select FAP$_0$ that
corresponds to $\rho_{\rm th}=8$ for the baseline all-sky search, the typical value used by LIGO searches
\cite{2013arXiv1304.0670L}, and the one considered above for the effect of template bank reduction. Note
that $\rho_{\rm th}$ is identical for all templates for the baseline search. We calculate the corresponding
$\rho_{\rm th}$ for externally triggered searches similarly to the analysis of the effect of reduced
template banks above. The results in Fig. \ref{figure:ROC} are shown such that FAP values are normalized
by FAP$_0$, and the expected detection rates are normalized by the detection rates $\mathcal{R}_{0}$
corresponding to those of the baseline searches. This allows for the quantification of the advantage of
utilizing the expected NS mass distribution in the test statistic.

The 'cutoff' model comparison represents the same improvement as the reduced-template-bank comparison above.
We find that the expected detection rate for the
cutoff model improves by $\sim 12\%$ and $35\%$ for the all-sky and externally triggered cases compared to the
baseline, respectively. For the 'weighted' model, we see improvements of $\sim 17\%$ and $68\%$ for the two cases, respectively.

\vspace{0.3cm}
\paragraph{Conclusion. ---}
We explored the increase in the detection rate of GWs from binary NS mergers
due to reducing the search template bank based on the expected mass distribution of NS in binaries.
We found an increase of $14\%$ for all-sky searches, and $61\%$ for an
externally triggered search. The higher increase for the externally triggered case was expected
since in that case the template bank is a more significant contributor to the search trial factor.
For both all-sky and externally triggered cases, the optimal detection strategy includes templates
that cover over $90\%$ of the expected binary NS mergers, therefore increasing the
detection rate can be achieved by covering a significant fraction of the binary NS
mass parameter space, i.e. hardly losing any astrophysical signal.

We constructed realistic search method that incorporates the expected binary NS mass
distribution in its test statistic, and calculated its sensitivity increase over using no
information about the mass distribution. We find that this realistic search achieves similar
sensitivity increases for both all-sky and externally triggered searches than what we found
by simply reducing the template bank.

Beyond the obvious benefit of increasing the detection rate, constraining the NS
template bank also significantly reduces the computational cost of the search, with the
required template bank being only $6\%$ and $3\%$ of the total template bank for the all-sky
and externally triggered cases, respectively.

Several future extensions of the present work will be interesting. Further observations of NS binary masses can help, e.g.
through establishing correlations between the two NS masses. While current masses are for nearby binaries, gaining farther
sample can improve detection prospects \cite{2014PhRvD..89f2002D}. Additionally, similar improvements can improve black hole-NS and
black hole-black hole mergers as well.

\begin{acknowledgments}
The authors thank Kipp Cannon, Tom Dent, John Veitch and Zsuzsa Marka for their useful suggestions. This paper has been approved for publication
by the LIGO Scientific Collaboration, with LIGO Document Number LIGO-P1500134.
IB and SM are thankful for the generous support of Columbia University in the City of New York and the National Science
Foundation under cooperative agreement PHY-1404462.
\end{acknowledgments}

\bibliographystyle{h-physrev}

\end{document}